\documentstyle[12pt,aps,fleqn]{revtex}
\input{psfig}

\newcommand{\beq}{\begin{eqnarray}}
\newcommand{\eeq}{\end{eqnarray}}
\begin{document}
\begin{center}
{\bf\LARGE Shape fluctuations of a Deforamable Body in a Randomly Stirred Host Fluid}\\\vspace{.7cm}
{\bf Gady Frenkel, Moshe Schwartz}\\
{\em Raymond and Beverly Sackler Faculty of Exact Sciences\\
School of Physics and Astronomy\\
Tel Aviv University, Ramat Aviv, 69978, Israel}
\end{center}
\begin{abstract}
Consider a deformable body immersed in an incompressible fluid
that is randomly stirred. Sticking to physical situations in which the
body departs only slightly from its spherical shape, we investigate the
deformations of the body. The shape is decomposed into spherical
harmonic modes. We study the correlations of these modes for a general
class of random 
flows that include the flow due to thermal agitation. Our results are
general in the sense that they are applicable to any body that is
described solely by the shape of its surface. 
\end{abstract}
%
\section{INTRODUCTION}
%
The current paper is the second in a series of papers that treat
systems of deformable objects immersed in an incompressible
liquid. 
These systems are common in nature ( milk, emulsions, composite fluids
)and have been studied extensively. Nevertheless, there are many
questions concerning them that need to be answered still. This is so, because the actual solution of
such systems is 
extremely difficult. The objects interact with themselves and with each
other via the hydrodynamic interactions, and form, in
general, a non-linear many-body problem. Moreover, each object has infinite
degrees of freedom that correspond to deformations, and all of them , of
all the objects, interact.
The problem is a little simplified if the deformation of the objects
from spherical shape remains small and the fluid is in the linear
regime, as the case of the Stokes approximation to the Navier -
Stokes equation.\\
Our final goal is to obtain the response of the composite system to a
given velocity field imposed on the liquid. The velocity field we have
in mind may be fixed in time like simple shear or randomly fluctuating
in time and space. Even in the first case the velocity field, each
object experience must have a random part due to the random passage of
other objects nearby. Therefore, effects of randomness are important
to the understanding of such systems.
We wish to know first, 
what is the response of a single deformable object to the random
velocity field, created by the other objects and the external
source. Once the response of a single object 
is known, we can obtain the response of the full, many body, system by
using the response of each body as a source of an additional velocity field.
In the former paper \cite{gady2} we studied the motion of the center of a
deformable object and derived the Mean Squared Displacement of its
center. 
In this paper we investigate the deformation degrees of freedom,
within the scope of the simplifications stated above, thus completing
the description of the response of a single object to a random
velocity field. 
Decomposing the deformation into spherical harmonic modes, we consider the
correlations between deformation modes. We find, among other things
that different modes are decoupled and that the correlation
function does not depend on the parameter $m$ of the $Y_{l,m}$
spherical harmonics (following from spherical symmetry). We also
obtain a method to calculate these correlations as a function
of time, build typical drop shapes from the correlation functions, and
discuss several interesting cases as thermal agitation.\\
The plan of this paper is as follows:
In section \ref{sec:system} we describe the system we have in mind and
formulate the basic equations. In section \ref{sec:white} we introduce
the deformation coefficients and obtain their correlations for the
simple case where the external velocity is uncorrelated in
time. General correlation functions are considered in section \ref{sec:general}. We obtain the deformation correlations and make
simplifications for equal times. In section \ref{sec:disscussion} we
discuss some properties of the correlation function of the external velocity
for several cases, among them is the special case of thermal
agitation. An algorithm for numerical computation of the correlations in the
general case is described in the Appendix.\\
%
%
\section{The system} \label{sec:system}
%
Consider a single deformable body immersed in a host fluid.
%
\begin{enumerate}
\item The deformable body is fluid, in the sense that the velocity
  fields are well defined everywhere (both inside and outside the body). In
  particular, each surface element moves with the velocity of the flow
  at its position: 
\beq  \label{eq:surfacebasic}
\dot{\vec{r}}=\vec{v}(\vec{r})
\eeq
\item  Both the body and the host fluid are incompressible,
$\vec{\nabla}\cdot\vec{v}=0$. 
\item The body is characterized by an energy that depends on its
shape ( Changing the orientation or switching places of two surface
particles while keeping the shape constant does not cause any
energy). The shape of minimum energy is a sphere. The
energy may be surface tension \cite{navot99}, Helfrich bending
energy \cite{helfrich76,lisy98}, etc.
Deformation of the shape change the energy, exerts force density in
the liquid and therefore creates an additional velocity field, denoted
$\vec{v}_{\psi}$.
\item We investigate the regime where the hydrodynamic equations are
  linear in the velocity ( i.e. a velocity field induced by several
  sources is equal to the linear sum of the velocity fields that each
  source induces separately). For instance, if the flow is governed by the
  Navier Stokes equation, then our assumption implies that the
  Reynolds number is small and that the stokes approximation is
  applicable. The linearity implies that in our system the actual
  velocity field is the linear sum of the imposed
  velocity field , $\vec{v}_{ext}$ ( where $\vec{v}_{ext}$ is
  the velocity field that would have been the actual velocity field if
  the body was absent), and the velocity field induced by the
  deformations, $\vec{v}_{\psi}$,
\beq
\vec{v} = \vec{v}_{ext} + \vec{v}_{\psi}.
\eeq
\item  We consider cases in which the external velocity field is
  random, with an average of zero. The velocity correlation function
  is known and depends only on distance and time difference. 
  We also assume that the the external velocity is weak
  enough to allow the body to remain almost spherical
%
%
%
%
\end{enumerate}
%
%
\section{White Noise Flow} \label{sec:white}
%
%
Consider a spherical body which is slightly deformed. The equation
\beq   \label{eq:fOmegat}
\frac{\rho}{R} + f(\Omega,t) - 1 = 0
\eeq
defines its
surface , providing for each spatial direction, $\Omega$, the distance,
$\rho \equiv |\vec{r} - \vec{r}_0|$, of the surface from the
center of the body, $\vec{r}_0$. $R$ is the radius of the undeformed sphere. $f(\Omega,t)$ that
parameterize the shape and is named the deformation function.
The deformation function is decomposed into spherical harmonics,
$f(\Omega,t)=\sum_{l=-\infty}^\infty \sum_{m=-l}^l
f_{lm}(t)Y_{lm}(\Omega)$, where the deformation coefficients are
denoted $f_{lm}(t)$. 
Our goal is to obtain
the correlations between deformation coefficients.\\
The center of the body, $\vec{r}_0$, is
chosen to be the point around which the deformation coefficients with
$l=1$ vanish: 
$f_{1,m}=0$. 
A different definition of the center will introduce three additional
equations for the deformation coefficients with $l=1$. We are not
interested in those since in the first order the spherical harmonics
with $l=1$ describe a translation of the body. In previous papers
\cite{gady2,gady-physica-A} we have used the same definitions to
describe the motion of the center. Therefore, their results are
consistent with the derivation done here and can be used.
%
Let $\psi(\vec{r})$ be a scalar field, defined everywhere in such a
way that the equation $\psi(\vec{r})=0$ describes the surface of the
body \cite{schwartz88,schwartz90b}. Straight forward manipulation of eq. (\ref{eq:surfacebasic})
give rise 
to a continuity equation for $\psi$, presented here in a coordinate
system that moves with the center of the body:
\beq   \label{eq:psi}
\dot{\psi} + \vec{v}_{\psi} \cdot \vec{\nabla} \psi = -(\vec{v}_{ext} -
\dot{\vec{r}}_0) \cdot \vec{\nabla} \psi
\eeq
%
A good candidate for $\psi$ is found in equation (\ref{eq:fOmegat}), $\psi =
\frac{\rho}{R} + f(\Omega, t) -1$. Assuming that $|\vec{v}_{ext} -
\dot{\vec{r}}|$ is small, the right hand side of eq. (\ref{eq:psi})
is, in the first order, equal to $Q \equiv \frac{1}{R}\left\{ \hat{\rho} \cdot
  [\vec{v}_{ext} -\dot{\vec{r}}]\right\}$ (see \cite{schwartz88}), where
$\hat{\rho}$ is a unit vector directed outwards from the center of the
body. Since the minimum energy of the body is obtained for a spherical shape,
the velocity induced by the body is zero when the sphere is
undeformed. Therefore, the velocity $\vec{v}_{\psi}$ must be, in
general, a linear functional of the deformation $f(\Omega,t)$
( with corrections of higher order that we neglect).
The term $\vec{v}_{\psi} \cdot \vec{\nabla} \psi$ is obtained in the
leading order by taking $\vec{\nabla} \psi$ on the original sphere and
$\vec{v}_{\psi}$ to first order in the deformation. We are interested
in the first order evolution of the surface, and the terms above are
already linear 
in the deformation; Thus, we can calculate them on the original
sphere, where it is possible to develop eq. (\ref{eq:psi}) in terms of
spherical harmonics (for further explanation see \cite{gady2}).
Consequently, the generic eq. for the deformation coefficient, $f_{l,m}$, must
be of the form
\beq \label{eq:basic}
\frac{\partial f_{lm}(t)}{\partial t} + \lambda_l f_{lm}(t) =
-Q_{lm}(t),
\eeq
where $Q_{lm}$ is given by
\beq   \label{eq:Qlm}
Q_{lm} 
= \frac{1}{R} \int d\Omega \left\{
    \hat{\rho}\cdot\left[\vec{v}_{ext} -
      \dot{\vec{r}}_0\right]Y^*_{l,m}(\Omega)\right\}
\eeq
and $\vec{v}_{ext}$ is evaluated on the surface of the body (and can
be approximated as on the undeformed body) at the
direction of the spatial angle $\Omega$.

It is convenient to write the correlation function of the external
velocity field in the momentum space. This is so because the random
velocity field is transversal when the fluid is
incompressible. Consequently, in real space, the flow must always be
correlated in a very complex way. On the other hand,
in the momentum space we can simply use a projection operator on the
transversal part of a general field:
%
\beq  \label{eq:vext}
\tilde{v}_{ext_i}(\vec{q}\ ) \equiv \sum_j \left( \delta_{ij}-\frac{q_i
    q_j}{q^2}\right) u_j(\vec{q}),
\eeq
where $\vec{u}$ is a general vector field and the bracketed term is the
projection operator that removes the longitudinal part of
$\vec{u}$, and therefore yields a general transversal velocity field $\tilde{v}_{ext}$.\\
Next, the correlations of the external velocity are easily expressed using the
correlations of the general field $\vec{u}$. We are interested in
cases were the system is isotropic, homogeneous and stationary. In
these cases, the general field must obey: 
\beq   \label{eq:fcorelation}
\left\langle u_l(\vec{q},t_1) u_m(\vec{p},t_2) \right\rangle =
\delta_{lm}\delta(\vec{q}+\vec{p})\phi(q , t_2-t_1),
\eeq 
where $\delta_{lm}$ is the Kronecker delta, $\delta()$ is the Dirac
delta function and $\phi()$ is a general function of $q$ and $\Delta
t$. In addition we have assumed that
\beq 
\left\langle u_l(\vec{q},t)\right\rangle = 0.
\eeq
In the rest of this section we consider a frequently
used family of random flows in which the external velocity is
uncorrelated in time.
\beq \label{eq:phiTilde}
\phi(q , t_2 - t_1) = \tilde{\phi}(q)\delta(t_2 - t_1)
\eeq
Equation (\ref{eq:phiTilde}) is a reasonable representative of
systems in which the smallest time scale is that of the random
velocity. In these systems we can replace the exact correlation
details by the effective delta function by integrating on time.\\
Using Fourier
transform and the definition of the correlations we can calculate the
correlation of the velocity at two places on the drop, that are
characterized by the directions $\hat{r}$ and $\hat{r'}$ . The calculation yields
\beq \label{eq:Vcorr1}
\left\langle v_{ext}^i(\hat{r},t_1)
v_{ext}^j(\hat{r}',t_2) \right\rangle = 
\int d\vec{q} e^{-i\vec{q}\cdot(\vec{r}-\vec{r}')}\left[\delta_{ij} -
  \frac{q_i q_j}{q^2}\right] \phi(|\vec{q}|)\delta(t_2 - t_1),
\eeq
where  $\vec{v}_{ext}(\hat{r},t)$ is the velocity at time $t$ at place
$\vec{r}$ on the surface and $\vec{r} \equiv \vec{r}_0(t)+R\hat{r}$. 
The average and correlations of $Q_{l,m}$ follow easily from the
previous equations.
The average of $Q_{lm}$ is zero,
\beq
\left\langle Q_{lm} \right\rangle = 0.
\eeq
%
The term $\hat{\rho} \cdot \dot{\vec{r}}_0$, in eq. (\ref{eq:Qlm}), does not
contribute to any component of $Q_{lm}$ except for those with
$l=1$. In addition, the center, $\vec{r}_0$, have been chosen to
be the point around which the three deformation coefficients,
$f_{lm}$, with $l=1$ are equal to zero. Therefore $Q_{1m}$ is
zero, and for all other choices $\dot{\vec{r}}_0$ can be dropped out
of $Q_{lm}$.\\ 
The external velocity on the surface is
approximated, in the first order of deformation, as the external
velocity on the undeformed sphere \cite{gady2}.
Strait forward calculation of $Q_{l,m}$ correlations using its
definition, eq. (\ref{eq:Qlm}), and the velocity correlations,
eqs. (\ref{eq:vext}-\ref{eq:phiTilde}), 
yields:
\beq  \label{eq:QlmShort}
\left\langle  Q_{lm}(t_1) Q_{l'm'}(t_2)\right\rangle =
\tilde{Q}_{lml'm'} \delta(t_2 -t_1) ,
\eeq
where
\beq   \label{eq:Q00}
\tilde{Q}_{lml'm'} \equiv
\frac{1}{R^2} \int d\Omega \int d\Omega' \int d^3q
Y_{lm}^*(\Omega)Y_{l'm'}^*(\Omega') \sum_{i,j =x,y,z} \Big[ \nonumber \\
\hat{r}_i \hat{r}_j' e^{-i\vec{q}\cdot(\hat{r} - \hat{r}')R}
\left[\delta_{ij} - \frac{q_i q_j}{q^2}\right] \phi(|q|) \Big].
\eeq
%
%
%
%

Finally, the correlations of the deformation coefficients, $f_{lm}$, are
obtained, using Eq. (\ref{eq:basic}) and the correlations of $Q_{lm}$, by direct integration
\beq
\left\langle  f_{lm}(t) f_{l'm'}(t+\Delta t)\right\rangle_{t
  \rightarrow \infty} = 
\lim_{t \rightarrow \infty}
\int_0^t dt_1 \int_0^{t+\Delta t} dt_2 \Bigg[ \nonumber \\
\left\langle  Q_{lm}(t_1)
  Q_{l'm'}(t_2)\right\rangle e^{\lambda_l (t_1 - t) + \lambda_{l'}(t_2
  - t- \Delta t)} \Bigg],
\eeq
where the limit $t \rightarrow \infty$ is taken to avoid the initial
conditions that are imposed on a specific realization.\\
The last equation combined with Eqs. (\ref{eq:Q00}) and
(\ref{eq:QlmShort}) yields
\beq  \label{eq:flmflmdelta}
\left\langle  f_{lm}(t) f_{l'm'}(t+\Delta t)\right\rangle_{t
  \rightarrow \infty} = 
\tilde{Q}_{lml'm'} \frac{e^{-
    \lambda_{l'}|\Delta t|}}{\lambda_l + \lambda_{l'}}.
\eeq
Since the system is invariant to time reversal it is obvious that there is
no preference of $\lambda_{l'}$ over $\lambda_{l}$ in the exponent, and
therefore the result must equal zero for $l \neq l'$. 
In fact, only for the terms in which $l=l'$ and $m = -m'$ $\left\langle
  Q_{lm}(0) Q_{l'm'}(0)\right\rangle \neq 0$. This follows from the
fact that in the first order $Q_{lm}$ depends only on the
component of the external velocity that is normal to the surface. Therefore,
its correlation function for two places at the same time must depend
only on the angle 
between their direction, and can be expanded as $\sum_l A_l
P_l(cos(\theta))$ where $P_l$ is the Legendre polynomial. This can be
turned into a sum of spherical harmonics using the partial waves expansion
\beq
P_l(cos(\theta)) = \frac{4 \pi}{2l +1} \sum_{m=-l}^{l} (-1)^m
Y_{lm}(\Omega)Y_{l\ -m}(\Omega')
\eeq
Now, it is obvious that the correlations of the deformation function,
$f(\Omega)$, must have the same
form
\beq
\left\langle f(\Omega)f(\Omega')\right\rangle = \sum_l B_l  \frac{4 \pi}{2l +1} \sum_{m=-l}^{l} (-1)^m
Y_{lm}(\Omega)Y_{l\ -m}(\Omega').
\eeq
Furthermore,
\beq
\left\langle f_{lm}f_{l'm'}\right\rangle =
\int \left\langle f(\Omega)f(\Omega')\right\rangle
Y_{lm}^*(\Omega)Y_{l'm'}^*(\Omega')\ d\Omega\ d\Omega' .
\eeq
Therefore, the expansion is composed of terms in which $l=l'$ and
$m=-m'$ only, as stated above.
\\
\\
%
%
\section{general noise} \label{sec:general}
%
%
In many cases white noise correlations are not sufficient to describe
what really happens in the liquid. Especially if the correlation time
is of the order of other time parameters. Such is the case of a system
of many droplets inside a host fluid. The random flow, a droplet is
subjected to, results from the random motion and deformation of other
droplets that pass nearby. It is obvious that in this case the
approximation of the flow to be uncorrelated in time is not
justified. Hence, we need to generalize the theory.

Suppose that $\phi(q,\Delta t)$ is a general function of $q$ and the
time differences $\Delta t$. The correlations
of the external velocity are now extended in time. In order to
calculate averages on the droplet in different times we must now
consider also the movement of the droplet. The definition of $Q_{l,m}$
(\ref{eq:Qlm}) 
implies that the correlations of the normal component of the external
velocity field on the surface of the body are
\beq \label{eq:QlmQlmGeneral}
\left\langle  Q_{lm}(t_1) Q_{l'm'}(t_2)\right\rangle =
\frac{1}{R^2} \int d\Omega \int d\Omega'
Y_{lm}^*(\Omega)Y_{l'm'}^*(\Omega') \sum_{i,j =x,y,z} \Big[ \nonumber \\
\hat{r}_i \hat{r}_j' \left\langle v_{ext}^i(\hat{r},t_1)
v_{ext}^j(\hat{r}',t_2) \right\rangle \Big].
\eeq
The correlation of
the velocity at two points on the droplet, located in the directions
$\hat{r}$ and $\hat{r}'$ and measured at different moments,  obviously
depends on the displacement of the center, $\Delta\vec{r}_0$. We 
calculate first the correlation for a general displacement of the center
and then, average the result according to the probability of finding
the center at each point.
To do this, we first express the velocity correlations in means of the
Fourier transform of the velocity, use equations
(\ref{eq:vext}),(\ref{eq:fcorelation}) and obtain
\beq
\left\langle v_{ext}^i(\hat{r},t_1)
v_{ext}^j(\hat{r}',t_2) \right\rangle =
\int P(\Delta \vec{r}_0) d(\Delta \vec{r}_0)
\int d^3q e^{-i\vec{q}\cdot(\Delta \vec{r}_0 + R(\hat{r} -
  \hat{r}')}[\delta_{ij}-\frac{q_i q_j}{q^2}]\phi(q,\Delta t), 
\eeq 
where $\Delta \vec{r}_0=\vec{r}_0(t_1) - \vec{r}_0(t_2)$, $\Delta t =
t_2-t_1$, $P(\Delta r_0)$ is the probability that the center will
be displaced by $\Delta \vec{r}_0$ in the period of $\Delta t$ and the
summation over 
$d(\Delta \vec{r}_0)$ is taken on all the possible configurations.\\
It is obvious now that averaging over the center displacements
will effect only the term $-i\vec{q} \cdot \Delta \vec{r}_0$ in the
  exponent since this is the only term that depends on $\Delta
 \vec{r}_0$. Consequently,
\beq
\left\langle v_{ext}^i(\hat{r},t_1)
v_{ext}^j(\hat{r}',t_2) \right\rangle =
\int d^3q \left\langle e^{-i\vec{q}\cdot \Delta \vec{r}_0}
\right\rangle e^{-i\vec{q}\cdot (\hat{r} -
  \hat{r}')R}(\delta_{ij}-\frac{q_i q_j}{q^2})\phi(q,\Delta t), 
\eeq
Assuming Gaussian distribution of the displacements of the center,
\beq
\left\langle e^{-i\vec{q}\cdot \Delta \vec{r}_0}
\right\rangle = e^{-\frac{q^2}{6}\langle (\Delta \vec{r}_0)^2\rangle}.
\eeq
In a previous paper \cite{gady2}, we considered the Mean Squared
Displacement (MSD) of the center of a
deformable body in a flow that is correlated in a general way. We
found that the MSD in a period of time $\Delta t$ is given by:
\beq \label{eq:MSDintegral}
F(\Delta t)=
16\pi \int_0^{\Delta t} dt' \int_0^\infty q^2dq
 e^{-\frac{q^2}{6}
F(t')}
 \phi (q,t')\left(j_0(qR)+j_2(qR)\right)^2(\Delta t-t'),
\eeq
where $F(\Delta t) \equiv \langle (\Delta \vec{r}_0)^2\rangle$.
Therefore the correlation of the external velocity at two points on
the surface, characterized by the directions $\hat{r}$ and $\hat{r}'$,
measured at two different times with time gap of $\Delta t$, is given
by
\beq \label{eq:FinalVelocityCorrelation}
\left\langle v_{ext}^i(\hat{r},t_1) v_{ext}^j(\hat{r}',t_2) \right\rangle =
\int d^3q  e^{-\frac{q^2}{6} F(\Delta t)}
e^{-i\vec{q}\cdot (\hat{r} -
  \hat{r}')R}(\delta_{ij}-\frac{q_i q_j}{q^2})\phi(q,\Delta t), 
\eeq
Equations (\ref{eq:MSDintegral}) and
(\ref{eq:FinalVelocityCorrelation}) enable us to calculate the
correlations of $Q_{l,m}$ (eq. (\ref{eq:QlmQlmGeneral})). The
correlations of the deformation coefficients can be obtained from their
basic equation (\ref{eq:basic}) using the correlations of $Q_{l,m}$,
\beq \label{eq:flmGeneralCorrelation}
\left\langle f_{l,m}(t) f_{l',m'}(t+\Delta t) \right\rangle_{t
  \rightarrow \infty}  =
\int_0^{t} dt' \int_0^{t+\Delta t} dt'' \frac{\left\langle Q_{l,m}(t') Q_{l',m'}(t'')
\right\rangle exp\left( \lambda_l t' + \lambda_{l'}
    t''\right)}{exp\left( \lambda_l t + \lambda_{l'} (t+\Delta t)\right)}.
\eeq
Equations (\ref{eq:QlmQlmGeneral}), (\ref{eq:MSDintegral}),
(\ref{eq:FinalVelocityCorrelation}) 
and (\ref{eq:flmGeneralCorrelation}) form the 
calculation method for the correlations of the
deformation coefficients. The method presented here may be hard to
implement to numerical calculations because of the many dimensions
integration. In the appendix we describe an algorithm that reduces the
integrations to be one dimensional, thus making the
computation task easier. This algorithm was used to obtain the results
presented in the following section.\\
At same times, $\Delta t=0$, the correlations above play an
important role in the 
shape distribution of the drop. They hold all the information needed
to determine the distribution of the $f_{l,m}$'s. As was explained in the
previous section, same time correlation (STC) are obviously non-zero
only for $l=l'$ and $m=-m'$. In addition we consider cases where
$\langle f_{l,m} \rangle=0$. Thus $\langle f_{l,m}(t) f_{l,-m}(t)
\rangle$  give the
variance of the deformation coefficient $f_{l,m}$, and since $f_{l,m}$
with different $l,m$ are uncorrelated 
we can use this variance to create typical shapes of drops
(Fig. \ref{fig:wnoise1}).\\
Equation (\ref{eq:flmGeneralCorrelation}) can be simplify a little in
the case of STC
\beq \label{eq:STC}
\left\langle f_{l,m}(t) f_{l',m'}(t) \right\rangle_{t
  \rightarrow \infty}  =
\int_0^{\infty} dt' \left\langle Q_{l,m}(0) Q_{l',m'}(t')
\right\rangle \frac{exp\left( -\lambda_l t'\right) + exp\left(- \lambda_{l'}
  t'\right)}{\lambda_l  + \lambda_{l'} }.
\eeq

%
%
\section{The Correlation function} \label{sec:disscussion}
%
%
We wanted the method presented above to be applicable to different
random external flows. Therefore, we have built the calculation method
in such a way that the correlation function $\phi(q,t)$ must be given
by hand. In some cases $\phi(q,t)$ can be calculated theoretically. In
others, to obtain $\phi(q,t)$ for a given system one can make an
experiment. Exclude the deformable body from the system; measure the
correlations in the velocity in different points at different times
and deduce the correlation function $\phi(q,t)$. To confirm the
results we present here just insert back the deformable body,
measure the correlations and compare with the 
theoretical prediction. We can also try to use our method to obtain
the correlation function itself. Of particular interest is the
response of a system of deformable bodies to an external flow. The
deformation of each body will induce an additional velocity field that
will influence the others. Hence, the correlations in the velocity,
each body is subjected to, can be treated in a self consistent manner,
using the deformations as a source to the external velocity field as
well as the field's product. 

In the rest of this section we will analyze examples of random flows.\\
%
%
Consider first the special case of thermal agitation. In a previous
paper \cite{gady2} we have shown that the correlation function for the external
velocity due to thermal agitation have the form $\phi(q,t)=\frac{K_B
  T}{(2 \pi)^3 \eta} \frac{\delta(t)}{q^2}$, where $\eta$ is the
viscosity of the fluid (notice that the correlations are long ranged
due to the incompressibility of the fluid). Using this form with
eq. (\ref{eq:flmflmdelta}) the STC of the
deformation coefficient, $f_{l,m}$, where $l>1$ is given by 
\beq \label{eq:thermal}
\left \langle f_{l,m}(t) f_{l,-m}(t) \right\rangle_{t \rightarrow \infty} = \frac{ K_B T}{R^2 \lambda} \frac{1}{(l+2)(l-1)}
\eeq

Schwartz and Edwards \cite{schwartz91} considered the special case of
deformable a body in equilibrium at temperature $T$, using the
Kirkwood equation. They found identical correlations in the
deformations (using $X_{l,m}=Rf_{l,m}$). Their derivation, however,
has been tailored for thermal agitation and 
cannot be expanded to take into account any other correlations in the
host fluid.

The shape of the deformable body is described by $r =
R(1+f(\Omega))$. Hence, eq. (\ref{eq:thermal}) implies that the
magnitude of the deformation, $Rf_{l,m}$, does not depend on the size
of the droplet, i.e. the deformations should be as of an infinite
lamellar surface.\\ 

\begin{figure}[!htp]
\centerline{\psfig{figure=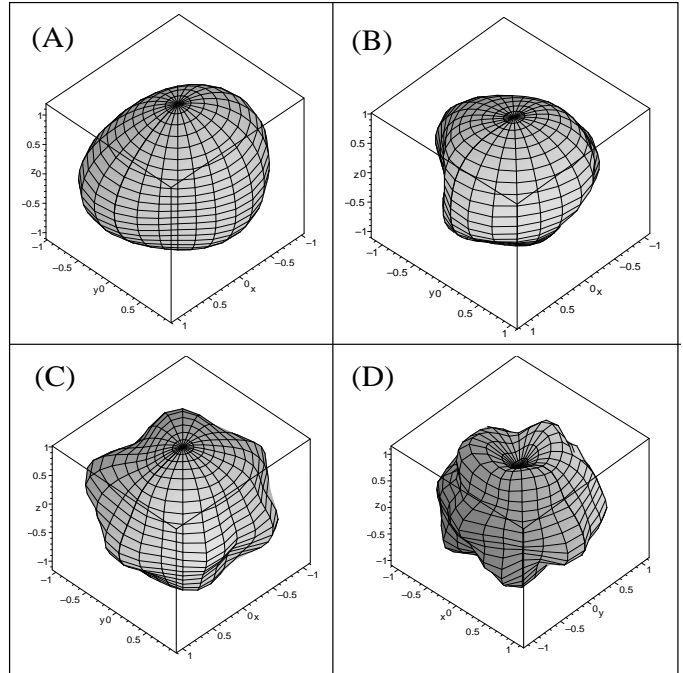,width=9cm ,height=9cm,clip=}}
\caption{A typical realization of a deformable body subjected to a
  random flow of the form $\phi(q)=C \delta(q\xi-1)\delta(t)$ with
  (A) $R=2\xi$, (B) $R=4\xi$, (C) $R=6\xi$ and (D) $R=8\xi$.}  

\label{fig:wnoise1}
\end{figure}

\begin{figure}[!htp]
\centerline{\psfig{figure=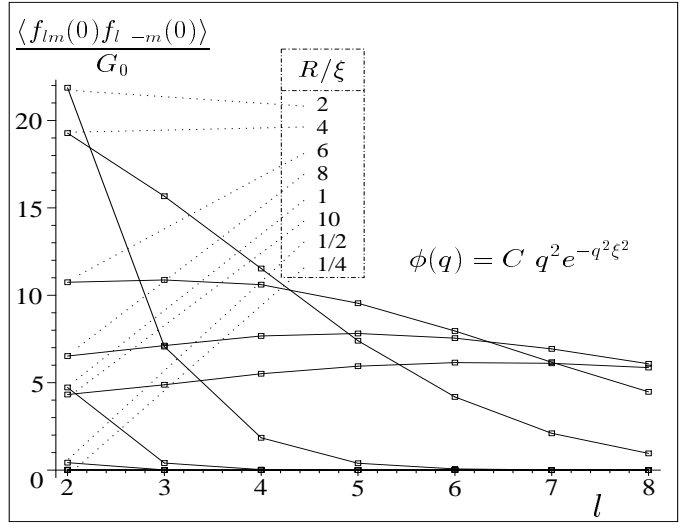,width=9cm ,height=7cm,clip=}}
\caption{The variance of the deformation coefficients, $f_{lm}$, as a
  function of $l$, for $\phi(q)=C\ q^2e^{-q^2\xi^2}$. ($G_0 \equiv \mu_2$).}  

\label{fig:flmflmq2exponent}
\end{figure}

\begin{figure}[!htp]
non-separable\centerline{\psfig{figure=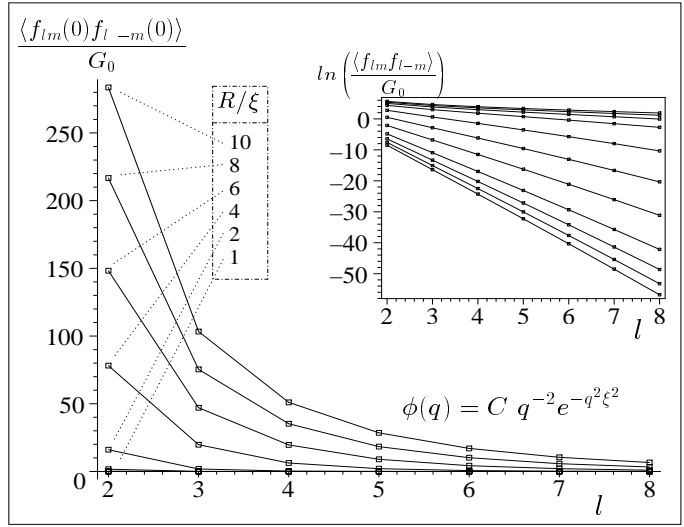,width=9cm ,height=7cm,clip=}}
\caption{The variance of the deformation coefficients, $f_{lm}$, as a
  function of $l$. $\phi(q)=Cq^{-2}e^{-q^2\xi^2}$. For $R>>\xi$ the
  curves coincide with the curves for
  thermal agitation. ($G_0 \equiv \mu_2$).}  

\label{fig:flmflmThermal}
\end{figure}

Thermal agitation belongs to a class of systems in which the decay
time, which is the time that takes the external velocity field to
loose memory and become uncorrelated, is the shortest time scale (Although
other time scales $\tau_l=\frac{1}{\lambda_l}$ tend to zero for $l
\rightarrow \infty$ we can use this form to any finite set of
spherical harmonics. In addition we must remember that $l$ is limited
by the fact that features on the sphere must be larger than the
inter-atomic distance for the continuum description to hold). In these
cases we can approximate the actual velocity field to be totally
uncorrelated in time.
\beq
\phi(q,t) = C \tilde{\phi}(q)\delta (t).
\eeq
There are two possible scenarios: either $\tilde{\phi}(q)$ decay with
a length scale, $\xi$ (at least one), or it is long-ranged. We limit
our discussion to one power and one length scale and the
generalization is trivial. In the first case $\tilde{\phi}(q) =
q^{-\alpha}\tilde{\tilde{\phi}}(\xi q)$, where $\tilde{\tilde{\phi}}$
has a cut off at $\xi q =1$. There are two dimensionless parameters for
the STC, $\mu_1=\frac{R}{\xi}$ and $\mu_2=\frac{C}{R^{5-\alpha}\lambda_l}$.
In the case of a body characterized by surface tension energy,
$\lambda_l \propto \frac{\lambda}{\eta R}$ where $\lambda$ is the
surface tension and $\eta$ is the viscosity, 
\beq 
\left\langle f_{l,m}(t) f_{l',m'}(t) \right\rangle_{t
  \rightarrow \infty}  =
F_{l,m,l',m'}\left( \frac{C \eta}{R^{4-\alpha} \lambda} ; \ 
  \frac{R}{\xi} \right).
\eeq
Fig. \ref{fig:flmflmq2exponent} and \ref{fig:flmflmThermal} depict the
STC dependence on $l$ for two correlation functions and different
parameters values where we changed
$\mu_1$ and $\mu_2$ by keeping $\xi=1$ and $\frac{C \eta}{\lambda}=1$
and changing $R$. As can be seen, there are two possibilities: either
the STC with $l=2$ dominates the curve or there is a maximum at $l_0
\approx \frac{R}{\xi}$. It is easy to see that the decay, for $l>>l_0$
is exponential. This suggests that there is a cutoff on the
deformation coefficients at which the expansion can be terminated
and therefore that the expansion we use here will be useful for
systems in which $\mu_1$ is small.\\
Fig \ref{fig:wnoise1} depicts various shapes of a body characterized
by surface tension under a random flow, given by
$\phi(q,t)=C\delta(\xi q - 1)\delta(t)$. As can be seen,
the surface of the body develops bumps. It is obvious that the
size of these surface features depends on the ratio $\mu_1=R/\xi$,
(Fig. \ref{fig:wnoise1}). As $\gamma$ increases,
different surface elements become less and less correlated.
Therefore, we expect to see features of smaller size (which correspond,
clearly, to spherical harmonics of  higher order). The smallest
features correspond to deformation coefficients with $l \simeq
\mu_1$ (or $l=2$ if $\mu_1 \leq 2$).\\
In the second case the parameter $\mu_1$ can be dropped out and we are
left only with $\mu_2$, as in the case of thermal agitation.\\
The correlation of the deformation coefficients for different times
$\left\langle f_{l,m}(t) f_{l',m'}(t+ \Delta t) \right\rangle_{t
  \rightarrow \infty}$ depend in addition on the dimensional parameter
$\mu_3=\lambda_l \Delta t$ exponentially as written in
eq. \ref{eq:flmflmdelta}. 
%
%
%
%
\begin{figure}[!htp]
\centerline{\psfig{figure=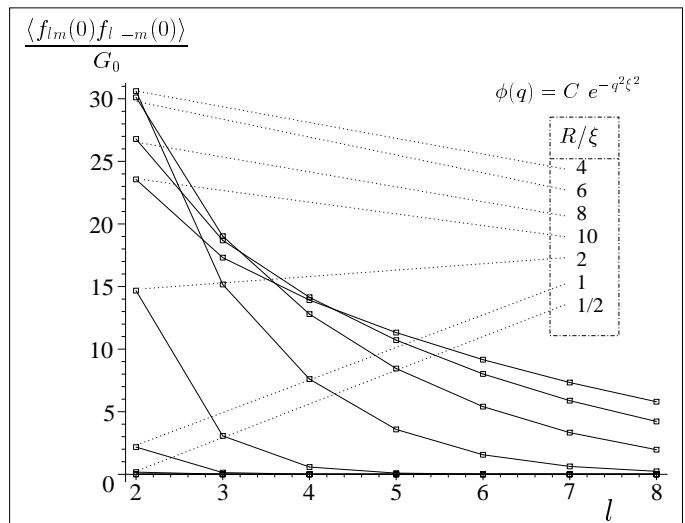,width=9cm ,height=7cm,clip=}}
\caption{The variance of the deformation coefficients, $f_{lm}$, as a
  function of $l$, for a typical decaying external noise :
  $\phi(q)=Ce^{-q^2\xi^2}$. ($G_0 \equiv \mu_2$).}  

\label{fig:flmflmexponent}
\end{figure}

Other classes of random flows include separable correlation functions
in which there is a decay length scale, $\xi$, and decay time scale
$\tau$: $\phi(q,t)=Cq^{-\alpha}\tilde{\phi}(\xi q)T(t/\tau)$, where 
$\left\langle f_{l,m}(t) f_{l',m'}(t) \right\rangle_{t \rightarrow \infty}  =
F_{l,m,l',m'}\left( \frac{C \tau^2}{R^{5-\alpha}} ; \ 
  \frac{R}{\xi}; \  \lambda_l\tau \right)$, and  flows of
the form $\phi(q,t)=Cq^{-\alpha}\tilde{\phi}(\Gamma q t^\beta)$ for
which analog dimensional parameters can be written.

\section{Summary} \label{sec:coclusions}
We have built a method to calculate the correlations of the
deformation coefficients $f_{l,m}$ that correspond to the decomposition
of the shape into spherical harmonics. We did it in two stages. The
first, for external velocity fields that are uncorrelated in time:
eqs. (\ref{eq:Q00}) , (\ref{eq:flmflmdelta}). The second, for a
general external velocity field that is correlated both in space and
time: eqs. (\ref{eq:QlmQlmGeneral}), (\ref{eq:MSDintegral}),
(\ref{eq:FinalVelocityCorrelation}) and
(\ref{eq:flmGeneralCorrelation}). The algorithm was based on the
assumption that the correlation of the velocity field in an identical
system without the body are known. We discussed these correlations,
used the results to build in a rigorous way typical surface shapes and
considered the special case of thermal agitation. In addition, we
pointed out that from our numerical results there there is a parameter
$\mu_1=R/\xi$ where deformation coefficients with $l>\mu_1$ seem to
decay exponentially and therefore are essentially unimportant. This
suggests that working with spherical harmonics to investigate systems
of deformable bodies is extremely useful for cases where $\mu_1$ is small.

\appendix
\section{algorithm for the calculation of the deformation correlations}
The velocity correlations involve a three dimensional integration, and
Eq. (\ref{eq:Q00}) uses an additional four dimensional integration.
As we can see, 
the method presented above is very hard to use. Therefore we must find
a way to lower the dimensionality of the integrals. the following
algorithm illustrates a method to do so using the addition theorem
(or partial waves expansion). The result is a finite expression which
is composed of sum of terms where each term involves only one
dimensional integration. Unfortunately, although finite, this sum is too
long to be presented here.\\
The partial waves expansion is given by
\begin{eqnarray}
e^{-i\vec{q}\cdot(R\hat{\rho})} = \sum_{l=0}^{\infty} \sum_{m=-l}^{l}
(-i)^l 4\pi j_l(qR)Y_{lm}^*(\Omega_q)Y_{lm}(\Omega) ,
\end{eqnarray}
where $\Omega_q$ is the solid angle in the $\vec{q}$ direction and 
$j_l$ is the spherical Bessel function.
Using the partial waves expansion with Eq. (\ref{eq:Vcorr1}) yields
\beq \label{eq:Vcorr2}
\left\langle v_{ext}^i(\vec{r}(t))
v_{ext}^j(\vec{r'}(t +\Delta t)) \right\rangle =
\sum_{l_1,l_2,m_1,m_2}
A_{l_1m_1l_2m_2ij}Y_{l_1m_1}(\Omega)Y_{l_2m_2}^*(\Omega') ,
\eeq
where 
\beq
A_{l_1m_1l_2m_2ij} \equiv& &(-i)^{l_1}\ i^{l_2}\ (4\pi)^2 \int d^3q\
e^{-\frac{q^2}{6}F(\Delta t)}  \ j_{l_1}(qR) j_{l_2}(qR)\cdot
\nonumber \\
& &Y_{l_1m_1}^*(\Omega_q)Y_{l_2m_2}(\Omega_q) \left[\delta_{ij} -
  \frac{q_i q_j}{q^2}\right] \phi(|\vec{q}|,\Delta t)
\eeq
Each coefficient $A_{lml'm'ij}$ divides easily into two parts, one
deals with the angular integration and the other with the radial
integration. The angular part is easily integrated, by the use of the
Clebsch-Gordan coefficients,  to give an algebraic
expression. Therefore to evaluate 
$A_{lml'm'ij}$ one needs to calculate the radial integration which is
one dimensional.\\
We use Eq. (\ref{eq:Vcorr2}) and obtain a new equation for the
$Q_{lm}$ correlations.
\beq
\left\langle  Q_{lm}(0) Q_{l'm'}(\Delta t)\right\rangle \equiv
\frac{1}{R^2} \int d\Omega \int d\Omega'
Y_{lm}^*(\Omega)Y_{l'm'}^*(\Omega') \sum_{i,j =x,y,z} \Big[ \nonumber \\
\hat{r}_i \hat{r}_j' \sum_{l_1,m_1,l_2,m_2} A_{l_1m_1l_2m_2ij}
Y_{l_2m_2}^*(\Omega') Y_{l_1m_1}(\Omega) \Big]
\eeq
The integrations over $\Omega$ and $\Omega'$ can be easily calculated once we
take into consideration that unit vectors can be decomposed into spherical harmonics
with $l = 0,1$. The integration yields a finite sum of
terms. Each term is composed by a known
number, $\zeta$, multiplied by the corresponding
$A_{l_1,m_1,l_2,m_2,ij}$ that involves only one dimensional integration.
Hence,
\beq
& &\left\langle  Q_{lm}(0) Q_{l'm'}(\Delta t)\right\rangle \equiv
\sum_{l_1,l_2,m_1,m_2,i,j} \nonumber\\ 
& &\zeta_{l_1,l_2,m_1,m_2,i,j}(l,l',m,m') \int q^2 dq
e^{-\frac{q^2}{6} F(\Delta t)} j_l(qR)
j_{l'}(qR)\phi(|\vec{q}|, \Delta t)
\eeq 
In this way we formulated a method for the calculation of the correlations.
The explicit expression is extremely long and therefore is not presented in
this paper.
\\
\\

\end{document}